\documentclass[12pt]{iopart}
\usepackage{graphicx}
\usepackage{hyperref}
\usepackage{epstopdf}
\newcommand{\beq}{\begin{equation}}
\newcommand{\eeq}{\end{equation}}
\newcommand{\bea}{\begin{eqnarray}}
\newcommand{\eea}{\end{eqnarray}}
\newcommand{\nn}{\nonumber}
\renewcommand{\d}{\partial}
\usepackage{iopams}
\begin{document}

\title[Instabilities near the onset of spin density wave order
in metals]{Instabilities near the onset of spin density wave order
in metals}

\author{Max A. Metlitski and Subir Sachdev}

\address{Department of Physics, Harvard University, Cambridge MA
02138, USA}
\ead{metlitski@physics.harvard.edu, sachdev@physics.harvard.edu}
\begin{abstract}
We discuss the low energy theory of two-dimensional metals near the onset of spin
density wave order. It is well known that such a metal has a superconducting instability induced
by the formation of spin-singlet pairs of electrons, with the pairing amplitude changing sign between regions of the Fermi surface
connected by the spin density wave ordering wavevector. 
Here we review recent arguments that there is an additional instability which is nearly as strong: towards the onset
of a modulated bond order which is locally an Ising-nematic order. This new instability is a consequence of an emergent `pseudospin' symmetry of the low energy
theory---the symmetry maps the sign-changing pairing amplitude to the bond order parameter.
\end{abstract}

\submitto{New Journal of Physics, special issue on ``Fermiology of Cuprates",\\ edited by Mike Norman and Cyril Proust}
\maketitle

\section{Introduction}

A number of recent experimental developments have refocused attention on a relatively simple picture \cite{scalapino, scalapino2}
of the origin of superconductivity in quasi-two-dimensional correlated electron compounds. 
Exchange of spin density wave (SDW) fluctuations induces an attractive interaction between electrons
in a spin-singlet, even-parity channel, with the pair wavefunction changing signs between regions of the
Fermi surface connected by the SDW ordering wavevector. 

In the pnictide superconductors, experiments identified
the nature of the SDW order and the configuration of the Fermi surface, and this theory then implies
superconductivity with $s_\pm$ pairing \cite{mazin,jiangping}, for which strong evidence has recently appeared \cite{hanaguri}.

In the cuprates, this picture correctly predicted the $d$-wave pairing signature.
However, it faced the difficulty that no strong SDW fluctuations were experimentally
observed in the optimal hole-doping region, where the superconductivity was the strongest.
A potential resolution is offered by the theory of competing orders \cite{sachdevzhang,zhang,qcnp}. In the strong-coupling regime,
the superconducting pairing amplitude has a significant 
feedback effect on the SDW fluctuations, and shrinks the region
of long-range SDW order \cite{moon}; consequently the SDW fluctuations are also substantially reduced at low temperatures
in the region of strongest superconductivity. An immediate consequence of this theory is that the
SDW order should re-emerge when superconductivity is suppressed by an applied magnetic field.
Such a re-emergence has been observed in a variety of experiments on both the LSCO and YBCO 
series of compounds \cite{lake,boris,chang1,chang2,mesot3}.
The high-field quantum oscillations \cite{louis1} are also naturally understood in this theory, 
with the re-emergent long-range SDW, or associated, order breaking up the Fermi surface into Fermi pockets. 

Motivated by these developments, we recently presented \cite{maxsdw} a detailed low energy description of the vicinity of the SDW transition
in two-dimensional metals. Here we will review our main results on the additional instabilities present near such a transition.
We will phrase our results for the case of the cuprate Fermi surface configuration, although there are natural extensions to 
the pnictides \cite{moon}.
As expected \cite{scalapino,ChubukovLong}, we find an instability to spin-singlet $d$-wave pairing. However, the pairing instability
is enhanced from the familiar BCS logarithmic divergence to a logarithm squared. We also find a dominant secondary instability
to a certain modulated bond order which is locally an Ising-nematic order: this instability is also associated with a susceptibility which diverges as a logarithm squared,
but with a co-efficient which is smaller than that of the pairing instability.

\section{Low energy theory}

We begin with a Hubbard-like model for electrons $c_{i \sigma}$ on the sites ($i$) of the 
square lattice ($\sigma = \uparrow\downarrow$ is the spin index) with dispersion $\varepsilon_{\vec{k}}$
as a function of wavevector $\vec{k}$. 
We decouple the repulsive interaction in the spin channel by a Hubbard-Stratonovich field $\phi^a$ ($a=x,y,z$).
This bosonic field $\phi^a$ represents collinear SDW order at wavevector $\vec{Q} = (\pi, \pi)$.
In this manner we obtain the familiar ``spin-fermion'' model for the vicinity of the
SDW ordering transition \cite{scalapino,ChubukovLong} with the Lagrangian
\bea
\mathcal{L} &=& \int_{\vec{k}} c^\dagger_{\vec{k} \sigma} \left( \frac{\partial}{\partial \tau} + \varepsilon_{\vec{k}} \right)
c_{\vec{k} \sigma} - \lambda \sum_i \phi^a_i c^{\dagger}_{i \sigma} \tau^a_{\sigma\sigma'} c_{i \sigma'}
e^{i \vec{Q} \cdot \vec{r}_i} \nn \\
&~& +  \int d^2 x \left[
\frac{1}{2 c^2} (\d_{\tau} {\phi}^a)^2 + \frac{1}{2}(\nabla_x {\phi}^a)^2 + \frac{r}{2} (\phi^a)^2 + \frac{u}{4} 
((\phi^a)^2)^2 \right] \label{spinfermion}
\eea
Here $\tau$ is imaginary time,  $\tau^a$ are the 
Pauli matrices, and we have partially integrated out high energy electrons to generate a bare
$\phi^4$ field theory Lagrangian for the SDW order parameter. 
The couplings $r$ and $u$ tune the strength of the SDW fluctuations: in mean-field theory there is an onset
of long-range SDW order at $r=0$.

The Lagrangian in (\ref{spinfermion}) is not yet
in the form of a low energy theory amenable to a continuum renormalization group (RG) analysis. This is because
of the transfer with the large wavevector $\vec{Q}$ in the spin-fermion term $\lambda$: this will scatter fermions
with momenta well away from the Fermi surface, and so involves high energy excitations.

\begin{figure}
\begin{center}
\includegraphics[width=3.5in]{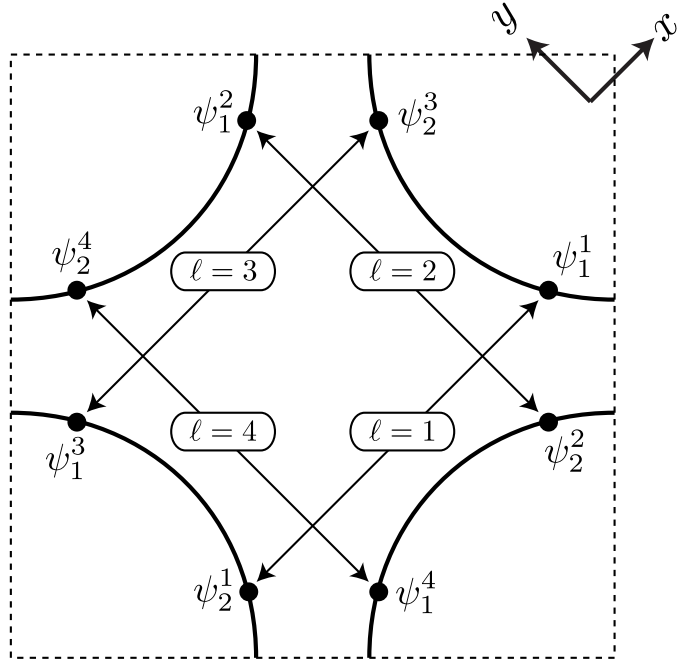}
\caption{Square lattice Brillouin zone showing the Fermi surface appropriate to the
cuprates. The filled circles are the hot spots connected by the SDW wavevector $\vec{Q} = (\pi,\pi)$.
The locations of the continuum fermion fields $\psi_1^\ell$ and $\psi_2^\ell$ are indicated.}
\label{fig:hotspots}
\end{center}
\end{figure}

\begin{figure}[t]
\begin{center}
\includegraphics[width=4in]{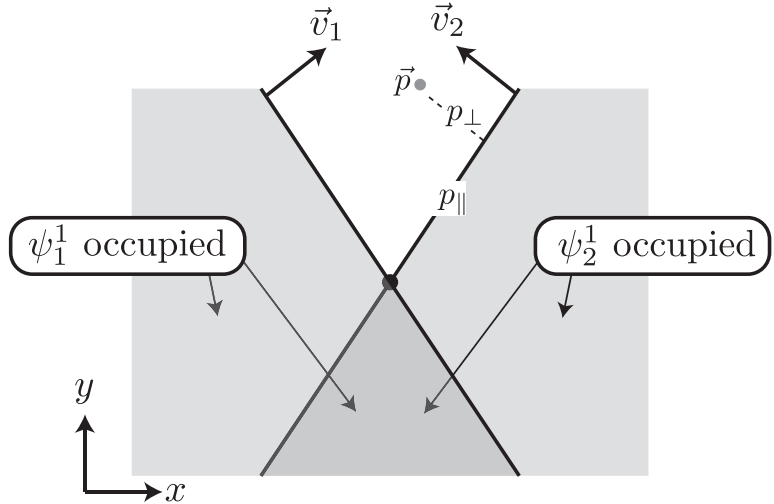}
\caption{Configuration of the $\ell=1$ pair of hot spots, with the momenta
of the fermion fields measured from the common hot spot at $\vec{k}=0$, indicated by the dark filled circle. The Fermi
velocities $\vec{v}_{1,2}$ of the $\psi_{1,2}$ fermions are indicated. The momentum components of the $\psi_2^1 (\vec{p})$ fermion parallel
($p_\parallel$) and orthogonal ($p_\perp$) to the Fermi surface are indicated.}
\label{fig:fermions}
\end{center}
\end{figure}

We focus on the low energy sector by zeroing in on the ``hot spots''. These are special locations on the 
Fermi surface where $\varepsilon_{\vec{k}+\vec{Q}} = \varepsilon_{\vec{k}}$: for these points, both the initial
and final electron states in the spin-fermion coupling can be right on the Fermi surface, and so the scattering
from the SDW fluctuations is the strongest.
For the electron dispersion appropriate to the cuprates, there are  $n = 4$ pairs of hot spots, as shown in Fig.~\ref{fig:hotspots}.
We introduce fermion fields $(\psi^{\ell}_{1\sigma}, \psi^{\ell}_{2\sigma})$, $\ell = 1 ... n$, $\sigma = \uparrow \downarrow$ for each pair of hot spots. Lattice rotations map the pairs of hot spots into each other, acting cyclically on the index $\ell$. Moreover, the two hot spots within each pair are related by a reflection across a lattice diagonal. It will be useful to promote each field $\psi$ to have $N$-flavors with an eye to performing a $1/N$ expansion. (Note that in Ref. \cite{ChubukovLong}, the total number of hot spots $2 n N$ is denoted as $N$.) The flavor index is suppressed in all the expressions. The low energy effective theory is then given by the following Lagrangian density in 2+1 
spacetime dimensions for the boson
$\phi^a$ and the fermions $\psi$
\bea L &=&   \frac{N}{2 c^2} (\d_{\tau} {\phi}^a)^2 + \frac{N}{2}(\nabla_x {\phi}^a)^2 + \frac{Nr}{2} (\phi^a)^2 + \frac{Nu}{4} 
((\phi^a)^2)^2\nn\\ &~&~~~~+
\psi^{\ell\dagger}_1 (\d_{\tau} - i \vec{v}^{\ell}_1 \cdot \nabla_x) \psi^{\ell}_1 + \psi^{\ell\dagger}_2 (\d_{\tau} - i \vec{v}^{\ell}_2 \cdot \nabla_x) \psi^{\ell}_2 \nn\\ &~&~~~~+ \lambda \phi^a \left(\psi^{\ell\dagger}_{1\sigma} \tau^a_{\sigma \sigma'} \psi^{\ell}_{2\sigma'} + \psi^{\ell\dagger}_{2\sigma} \tau^a_{\sigma \sigma'} \psi^{\ell}_{1\sigma'}\right). \label{L}\eea
The first line in Eq.~(\ref{L}) is the usual O(3) model for the SDW order parameter, the second line is the fermion kinetic energy, and the third line is the interaction between the SDW order parameter and the fermions at the hot spots. Here, we have linearized the fermion dispersion near the hot spots, and $\vec{v}^{\ell}$ are the corresponding Fermi velocities. It is convenient to choose coordinate axes along directions $\hat{x} = \frac{1}{\sqrt{2}}(1,1)$ and $\hat{y} = \frac{1}{\sqrt{2}} (-1,1)$, so that 
\begin{equation}
\vec{v}^{\ell = 1}_1 = (v_x, v_y)~~,~~\vec{v}^{\ell = 1}_2 = (-v_x, v_y),\label{vell1}
\end{equation}
The velocity ratio $\alpha = v_y/v_x$ plays an important role in the RG analysis of (\ref{L}).
The Fermi velocities (\ref{vell1}) are indicated in Fig.~\ref{fig:fermions}.
The other Fermi velocities are related by rotations, $\vec{v}^{\ell} = (R_{\pi/2})^{\ell-1} \vec{v}^{\ell = 1}$.

\section{Pseudospin symmetry}
\label{sec:pseudospin}

We now note a crucial symmetry of the theory (\ref{L}), which will be important for the instabilities discussed in 
this paper.
Besides the microscopic translation, point-group, spin-rotation and time-reversal symmetries, the low energy theory possesses a set of four emergent SU(2) pseudospin symmetries associated with 
particle-hole transformations. Let us introduce a four-component spinor,
\beq \Psi^{\ell}_i = \left(\begin{array}{c} \psi^{\ell}_i\\ i \tau^2 \psi^{\ell \dagger}_i\end{array}\right)\label{Psi}\eeq
We will denote the particle-hole indices in the four-component spinor by $\alpha, \beta$. The spinor (\ref{Psi}) satisfies the hermiticity condition,
\beq i \tau^2 \left(\begin{array}{cc} 0 & -1\\ 1 & 0\end{array}\right) \Psi^{\ell}_i = \Psi^{\ell \ast}_i \label{Herm}\eeq
Then, the fermion part of the Lagrangian (\ref{L}) can be rewritten as,
\bea L_\psi &=& \frac{1}{2} \Psi^{\ell\dagger}_1 (\d_{\tau} - i \vec{v}^{\ell}_1 \cdot \nabla) \Psi^{\ell}_1 + \frac{1}{2}\Psi^{\ell\dagger}_2 (\d_{\tau} - i \vec{v}^{\ell}_2 \cdot \nabla) \Psi^{\ell}_2  \nn \\
&~&~~~~~~+ \frac{1}{2}\lambda {\phi}^a \cdot \left(\Psi^{\ell\dagger}_{1} {\tau}^a \Psi^{\ell}_{2} + \Psi^{\ell\dagger}_{2} {\tau}^a \Psi^{\ell}_{1}\right). \label{LPsi}\eea
Now the Lagrangian (\ref{LPsi}) and the condition (\ref{Herm}) are manifestly invariant under,
\beq {\rm SU}(2)_{\ell}:\,\, \Psi^{\ell}_i \to U_{\ell} \Psi^{\ell}_i \quad , \quad \phi^a \rightarrow \phi^a \label{SU2}\eeq
where the $U_{\ell}$ are SU(2) matrices. Note that there are 4 independent SU(2) pseudospin symmetries, one for
each pair of hot spots. 
The diagonal subgroup of (\ref{SU2}) is associated with independent conservation of the fermion number at each hot spot pair. 

The symmetry (\ref{SU2}) is a consequence of linearization of the fermion spectrum near the hot spots and is broken by higher order terms in the dispersion. The diagonal subgroup noted above is preserved by higher order terms in the dispersion, but is broken by four-fermion interactions, which map fermion pairs from opposite hot spots into each other. Both symmetry breaking effects are irrelevant \cite{maxsdw} in the low energy limit used to derive (\ref{L}).

This large quadrupled SU(2) symmetry is 
thus a generic feature of the vicinity of the SDW ordering transition in metals.
It should be contrasted with the more familiar lattice SU(2) pseudospin symmetry of the Hubbard model \cite{yang,hkee}:
there is only a single such SU(2) symmetry,
present at half-filling, and only if the dispersion has a particle-hole symmetric form {\em i.e.\/}
the electron hopping is always between opposite sublattices. No such restrictions are placed in our case.
 We will also find a connection 
between instabilities in the particle-hole and particle-particle channels, but both will apply to the generic
spin-fermion model in (\ref{L}).

The pseudospin symmetry (\ref{SU2}) constrains the form of the fermion Green's function to be,
\beq - \langle \Psi^{\ell}_{i \alpha \sigma} \Psi^{m \dagger}_{j \beta \sigma'}\rangle = \delta^{\ell m} \delta_{ij}\delta_{\alpha \beta} \delta_{\sigma \sigma'}  G^{\ell}_i(x-x')\eeq
which implies,
\beq G^{\ell}_i(x-x') = - G^{\ell}_i(x'-x) \eeq
The corresponding  expression in momentum space, $G^{\ell}_i(k) = - G^{\ell}_i(-k)$, implies that the location of hot spots in the Brillouin zone is not renormalized by the spin wave fluctuations in the low energy theory.

\section{Pairing instability}
\label{sec:pairing}

\begin{figure}
\begin{center}
\includegraphics[width=3in]{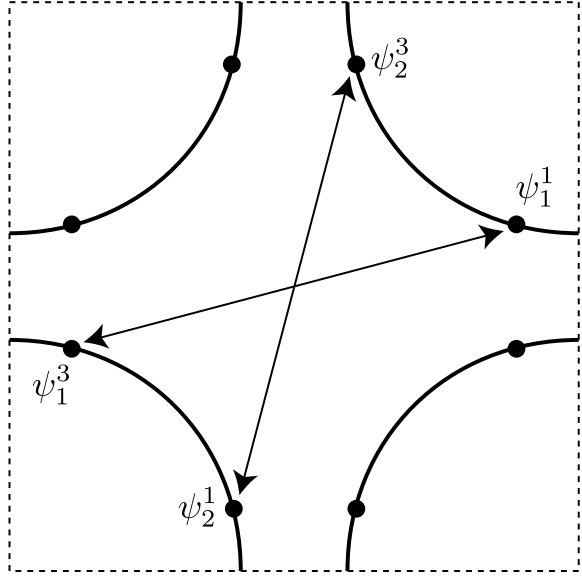}
\caption{Pairing of the electrons at the $\ell=1,3$ hot spots of Fig.~\ref{fig:hotspots}.
Electrons at opposite ends of the arrows form spin-singlet pairs.
The pairing amplitude is in the spin-singlet channel and has opposite signs
on the two arrows. There is a similar pairing instability between the $\ell=2,4$ hot spots.
}
\label{fig:hotspots2}
\end{center}
\end{figure}
A detailed RG analysis of the SDW transition described by (\ref{L}) was presented in Ref.~\cite{maxsdw} using the
framework of the $1/N$ expansion. We will not describe this here, but will use a few key results below.

Let us directly proceed to discuss the nature of the pairing instability. The pairing is induced by $\phi$-fluctuations,
which play the role of the phonon mode in conventional BCS theory. The pairing involves 2 pairs of hot spots, consisting
of time-reversed pairs of electrons, as shown in Fig.~\ref{fig:hotspots2}. The spin singlet superconducting order parameter, associated with Fig.~\ref{fig:hotspots2} is
\beq C = \epsilon_{\sigma \sigma'} (\psi^{1\dagger}_{1 \sigma} \psi^{3\dagger}_{1 \sigma'} -  \psi^{1\dagger}_{2 \sigma} \psi^{3\dagger}_{2 \sigma'}) \label{Cooperon}\eeq
and the relative sign between the two terms is the signature of $d$-wave pairing. Thus, the operator (\ref{Cooperon}) is odd under a reflection about the lattice diagonal $P : (k_x, k_y) \to (-k_x, k_y)$. 

The pairing instability in the channel (\ref{Cooperon}) is described by computing the four-fermion scattering amplitude,
\bea &&\Gamma_{{\rm BCS}}(\vec{k},\omega_1,-\vec{k},\omega_3;\vec{k}',\omega'_1,-\vec{k}',\omega'_3) = \nn\\
&&~~~\epsilon_{\sigma \sigma'}\epsilon_{\rho \rho'} \Biggl\langle \Bigl(\psi^{1}_{1 \sigma}(\vec{k'},\omega'_1) \psi^{3}_{1 \sigma'}(-\vec{k}',\omega'_3) - \psi^{1}_{2 \sigma}(P \vec{k}', \omega'_1) \psi^{3}_{2 \sigma'}(-P \vec{k}',\omega'_3)\Bigr)\nn\\&&~~~\times \Bigl(\psi^{3 \dagger}_{1 \rho'}(-\vec{k},\omega_3) \psi^{1\dagger}_{1 \rho}(\vec{k},\omega_1) - \psi^{3\dagger}_{2 \rho'}(- P \vec{k}, \omega_3) \psi^{1\dagger}_{2 \rho}(P \vec{k}, \omega_1)\Bigr)\Biggr\rangle,
\label{GammaBCS}\eea
where it is understood that the external fermion Green's functions have been truncated.
\begin{figure}
\begin{center}
\includegraphics[width=3.5in]{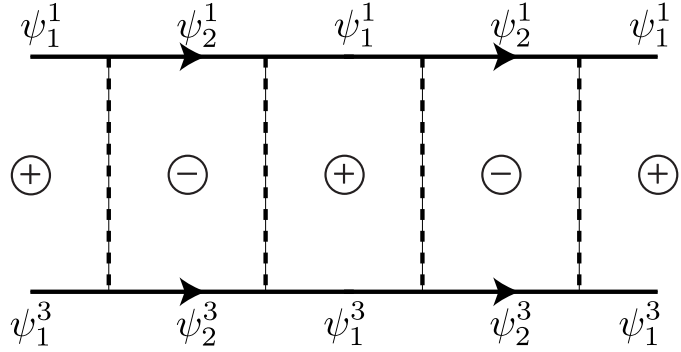}
\caption{Cooperon diagram associated with the pairing instability of the $\ell=1,3$ hot spots. 
The full lines are the fermion propagators, as indicated, and the dashed line is the propagator
of the SDW fluctuation $\phi^a$. The circled signs indicate the signs of the pairing amplitude of the corresponding
pairs of fermions.}
\label{fig:cooperon}
\end{center}
\end{figure}
As in conventional BCS theory, we will study the contribution of the Cooperon ladder diagrams illustrated in Fig.~\ref{fig:cooperon} to the amplitude (\ref{GammaBCS}). Here we will only examine the first two terms in the series, as shown in Fig.~\ref{fig:cooperon2}. 
\begin{figure}
\begin{center}
\includegraphics[width=4.5in]{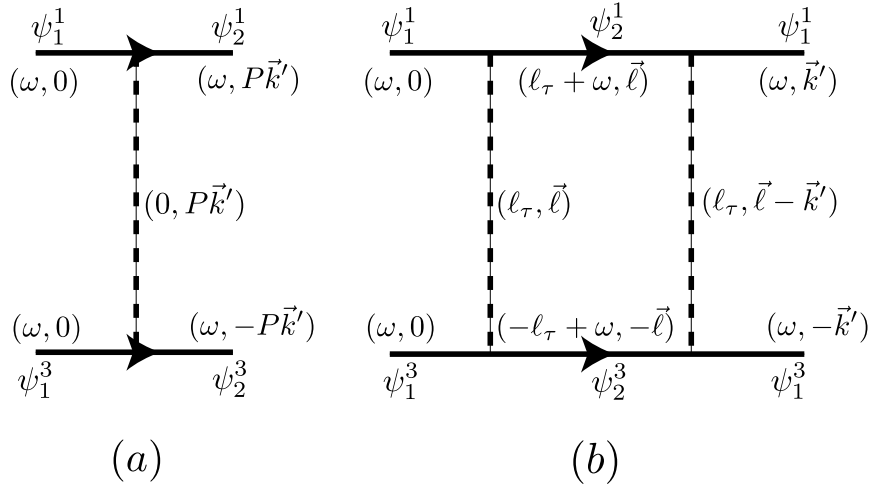}
\caption{The first two terms in the ladder series of Fig.~\ref{fig:cooperon}, with frequency/momentum labels,
as evaluated in Eqs.~(\ref{BCStree}) and (\ref{BCS1loop}).}
\label{fig:cooperon2}
\end{center}
\end{figure}
The dashed line denotes the propagator of the $\phi$ field, representing the SDW fluctuations. For this we use the overdamped paramagnon form appropriate to the $1/N$ expansion,
\beq D(\omega,\vec{q}) = \frac{1}{N} \frac{1}{(\gamma |\omega| + \vec{q}^2)},\label{Dq}\eeq
where $\gamma = n\lambda^2/(2 \pi v_x v_y)$ is the damping constant obtained from a decay of a $\vec{\phi}$ into a particle-hole pair. 

The strongest infra-red enhancement of the scattering amplitude by SDW fluctuations occurs in the kinematic regime when the incoming fermions have momenta at hot spots $\vec{k} = 0$, while the outgoing fermions are ``cold", {\it i.e.} their momentum $\vec{k}'$ is along the Fermi surface and their frequency satisfies $\gamma |\omega| \ll \vec{k}'^2$. For simplicity, we set all the external frequencies to be equal $\omega_1 = \omega_3 = \omega'_1 = \omega'_3 = \omega$. With the above choice of external momenta the diagram in Fig.~\ref{fig:cooperon2}a evaluates to
\beq \Gamma^{\rm tree}_{{\rm BCS}} = \frac{12 \lambda^2}{N \gamma |\vec{k}'|^2}\label{BCStree}\eeq 

Next, we consider the diagram for the BCS amplitude in Fig.~\ref{fig:cooperon2}b, which evaluates to
\bea \Gamma^{\rm 1 loop}_{{\rm BCS}} = 36 \lambda^4 \int \frac{d \ell_\tau d^2 \vec{\ell}}{(2 \pi)^3} D(\ell_\tau, \vec{\ell})G^1_2(\ell_\tau+\omega,\vec{\ell}) G^3_2(-\ell_\tau + \omega,-\vec{\ell}) D(\ell_\tau,\vec{\ell}-\vec{k}') \nn \\
\label{BCS1loop}
\eea
where $G$ denotes the fermion Green's function. The main contribution to Eq.~(\ref{BCS1loop}) comes from momenta $|\vec{\ell}| \ll |\vec{k}'|$, $\gamma |\ell_\tau| \ll \vec{k}'^2$ such that the first scattering in Fig.~\ref{fig:cooperon2}b is ``soft", while the second one is ``hard". Then,
\beq \Gamma^{\rm 1 loop}_{{\rm BCS}} \approx \frac{36 \lambda^4}{N \gamma |\vec{k}'^2|} \int  \frac{d \ell_\tau d^2 \vec{\ell}}{(2 \pi)^3} D(\ell_\tau, \vec{\ell})G^1_2(\ell_\tau+\omega,\vec{\ell}) G^3_2(-\ell_\tau + \omega,-\vec{\ell}) \label{BCS1loop2}\eeq
Now we need the form of $G$ near the SDW critical point. This was described in some detail in Ref.~\cite{maxsdw} to one-loop order in the $1/N$ expansion: it has a complex structure depending upon the value of frequency and momentum from the hot spot. For the BCS scattering amplitude, it turns out \cite{maxsdw} that the dominant contribution comes from the Fermi liquid poles on the ``cold lines" in the vicinity of the hot spot. Let us define $p_\perp$ as the distance to the Fermi surface, and $p_\parallel$ as the distance to the hot spot (see Fig.~\ref{fig:fermions});  then there are well-defined Landau quasiparticles 
for $p_\perp \ll p_\parallel$ and $\gamma |\omega| \ll p^2_\parallel$, with
\beq G(\omega, \vec{p}) \sim \frac{{\cal Z} (p_\parallel )}{ i \omega - v_F (p_\parallel) p_\perp} .\label{GFSscal}\eeq
The Fermi velocity $v_F$ and the quasiparticle residue ${\cal Z}$ both vanish
linearly with $p_\parallel$ as we approach the hot spot 
\cite{maxsdw},
\beq v_F (p_\parallel ) = \frac{4 n N}{3\gamma \lambda^2} p_\parallel, \quad {\cal Z} (p_\parallel ) = \frac{4 N}{3\lambda^2} (2 \pi n)^{1/2} \gamma^{-1/2} \left(\frac{1}{\alpha} + \alpha\right)^{-1/2} p_\parallel, \label{ZVoneloop}\eeq
where recall $\alpha = v_y/v_x$. We now insert (\ref{Dq}), (\ref{GFSscal}), and (\ref{ZVoneloop}) into (\ref{BCS1loop2}) and obtain
\bea
\Gamma^{\rm 1 loop}_{{\rm BCS}} &=& -\frac{36 \lambda^4}{N^2 \gamma |\vec{k}'^2|}\int \frac{d\ell_\parallel}{2 \pi} \int \frac{d \ell_\tau}{2 \pi} \int \frac{d\ell_\perp}{2\pi} \frac{1}{(\gamma |\ell_\tau| + \ell^2_\parallel)}  \nn \\
&~&~~~~~~~~~~~\times  \frac{{\cal Z}(\ell_\parallel)}{\left[ i (\ell_\tau + \omega) - v_F(\ell_\parallel) \ell_\perp \right]} 
\frac{{\cal Z}(\ell_\parallel)}{\left[ i (\ell_\tau- \omega) + v_F(\ell_\parallel) \ell_\perp \right]},
\label{GammaBCS3}\eea
where $\ell_{\parallel,\perp}$ are defined with respect to the $\psi_{2}^{1,3}$ Fermi surfaces, as in Fig.~\ref{fig:fermions}.
Recalling that the internal fermions are taken to be cold, {\it i.e.} $\gamma |\ell_\tau| \ll \ell^2_\parallel$, we may approximate the bosonic propagator in Eq.~(\ref{GammaBCS3}) by its static value. After changing variables to $\epsilon = v_F(\ell_\parallel) \ell_\perp$ we then obtain,
\bea 
\Gamma^{\rm 1 loop}_{{\rm BCS}}  &=& -\frac{36 \lambda^4}{N^2 \gamma |\vec{k}'^2|} \int \frac{d\ell_\parallel}{2 \pi} \frac{{\cal Z}^2(\ell_\parallel)}{ v_F(\ell_\parallel)\ell^2_\parallel } \nn \\
&~&~~~~~~~\times \int_{\gamma |\ell_\tau| \lesssim \ell^2_\parallel} \frac{d \ell_\tau}{2 \pi} \int \frac{d\epsilon}{2\pi} \frac{1}{\left[ i (\ell_\tau + \omega) - \epsilon \right]} \frac{1}{\left[ i (\ell_\tau- \omega) + \epsilon \right]}. 
\eea
The integral over $\ell_\tau$, $\epsilon$ has the form familiar from Fermi-liquid theory and gives the usual BCS logarithm,
\beq \int \frac{d \ell_\tau}{2 \pi} \int \frac{d\epsilon}{2\pi} \frac{1}{\left[ i (\ell_\tau + \omega) - \epsilon \right]} 
\frac{1}{\left[ i (\ell_\tau- \omega) + \epsilon \right]} = - \frac{1}{2 \pi} \log \frac{\Lambda_{FL}}{|\omega|} \eeq
where $\Lambda_{FL}$ is the frequency/energy cut-off, which in the present case is $\Lambda_{FL} \sim \ell^2_\parallel/\gamma$. Of course, for the above form to hold, we need $|\omega| \ll \Lambda_{FL}$. Thus,
\bea \Gamma^{\rm 1 loop}_{{\rm BCS}}  &=& \left(\frac{12 \lambda^2}{N \gamma |\vec{k}'|^2}\right)  \frac{3 \lambda^2}{2 \pi^2 N} \int_{\sqrt{\gamma \omega}}^{|\vec{k}'|} d\ell_\parallel \frac{{\cal Z}^2(\ell_\parallel)}{v_F(\ell_\parallel)\ell^2_\parallel } \log \frac{\ell^2_\parallel}{\gamma |\omega|} \nn
\\ &=& \left(\frac{12 \lambda^2}{N \gamma |\vec{k}'|^2}\right) \frac{\alpha}{\pi (\alpha^2 + 1)} \log^2 \frac{\vec{k}'^2}{\gamma |\omega|} \label{logsq} \eea
where we have cut-off the ultra-violet divergence of the $\ell_\parallel$ integral by $|\vec{k}'|$. Comparing
Eq.~(\ref{logsq}) with Eq.~(\ref{BCStree}) we see that there is an enhancement of the Cooperon propagator
by the factor
\beq
1 + \frac{\alpha}{\pi (\alpha^2 + 1)} \log^2 \frac{\vec{k}'^2}{\gamma |\omega|} \label{fac1}
\eeq
which has the promised log-squared form. Note that this is not suppressed by a factor of $1/N$.
As we noted, the critical SDW
fluctuations enhance the BCS logarithm to the stronger divergence above. Note that this divergence came
from ``cold'' internal fermion lines in Fig.~\ref{fig:cooperon2}b; thus, similar effects will also be present
in higher order graphs in the Cooperon ladder.

It is not clear how to improve Eq.~(\ref{fac1}) using the RG.
However, we can note that the coupling $\alpha$ is of order unity, and so the pairing is enhanced as the frequency crosses the Fermi
energy.
We also note that in the two-loop RG, the 
coupling $\alpha = v_y/v_x$ has a flow towards weak coupling
\begin{displaymath}
\frac{d \alpha}{d \ell} = - \frac{12}{\pi n N} \frac{\alpha^2}{(1 + \alpha^2)}
\end{displaymath}
but it not appropriate to simply insert the integrated value from this flow into the pairing enhancement.

\section{Bond order instability}
\label{sec:nematic}

\begin{figure}
\begin{center}
\includegraphics[width=3.5in]{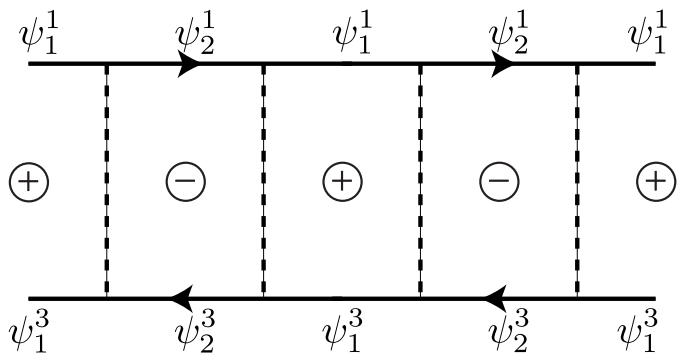}
\caption{Pseudospin partner of the Cooperon susceptibility in Fig~\ref{fig:cooperon}, obtained by 
reversing the line of the $\ell = 3$ fermion. As described in the text, this represents the susceptibility
to a modulated bond order which is locally an Ising-nematic order.
}
\label{fig:nematic}
\end{center}
\end{figure}

Let us now apply the pseudospin transformation in (\ref{SU2}) to {\em only\/} the $\ell =3$ pair of hot spots,
while leaving the $\ell=1$ pair unchanged. This will transform the Cooperon instability in the particle-particle
channel to an instability in the particle-hole channel. Diagrammatically, this corresponds to reversing
the $\ell=3$ line in Fig.~\ref{fig:cooperon} to obtain the susceptibility in Fig.~\ref{fig:nematic}.
By the symmetry of the low energy theory in Section~\ref{sec:pseudospin}, it would seem we can immediately
conclude that the scattering amplitude in Fig.~\ref{fig:nematic} should also have a log-squared
divergence at one loop order as in (\ref{logsq}). However, as was shown in \cite{maxsdw}, the naively irrelevant curvature of the Fermi surface, which breaks the pseudospin symmetry, modifies the result in the particle-hole channel at very low frequencies $|\omega| \ll A |\vec{k}'|^3$, where $A$ is a co-efficient related to the curvature. 
It was found that in this regime, the log-squared divergence did indeed survive, but with a co-efficient which was
smaller by a factor of 3. Thus in place of the factor in Eq.~(\ref{fac1}) for Fig.~\ref{fig:cooperon},
the amplitude in Fig.~\ref{fig:nematic} has the enhancement factor
\beq
1 + \frac{\alpha}{3\pi (\alpha^2 + 1)} \log^2 \frac{\vec{k}'^2}{\gamma |\omega|}. \label{fac2}
\eeq
Moreover, for higher frequencies, $\omega \gg A |\vec{k}'|^3$, the particle-particle and particle-hole channels are degenerate.

\begin{figure}[h]
\begin{center}
\includegraphics[width=3.3in]{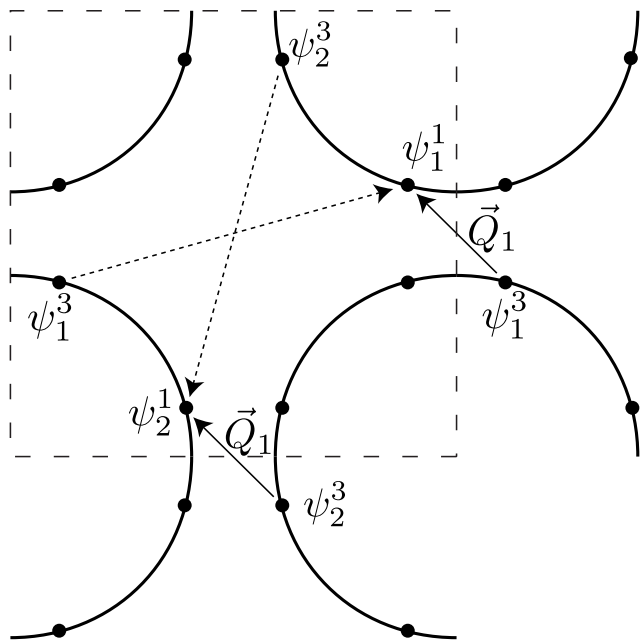}
\caption{Spin singlet density operators ($\sim \psi^\dagger \psi$)
of the electrons at the $\ell=1,3$ hot spots of Fig.~\ref{fig:hotspots} 
shown with an arrow pointing
from the Brillouin zone location of $\psi$ to that of $\psi^\dagger$. The dashed arrows are
the density operators in the first Brillouin zone. The full arrows are in an extended zone scheme
which shows that these operators have net momentum
$\vec{Q}_1 = 2 K_y (-1,1)$, where $(K_x, K_y)$ is the location of the $\ell=1$, $i=1$ hot spot (we have $K_x + K_y = \pi$).
The density operator with opposite signs on the two arrows is enhanced near 
the SDW critical point. Similarly the $\ell = 2,4$ hot spots contribute density operators
at $\vec{Q}_2 = 2 K_y (1,1)$.
}
\label{fig:hotspots3}
\end{center}
\end{figure}

\begin{figure}
\begin{center}
\includegraphics[width=4in]{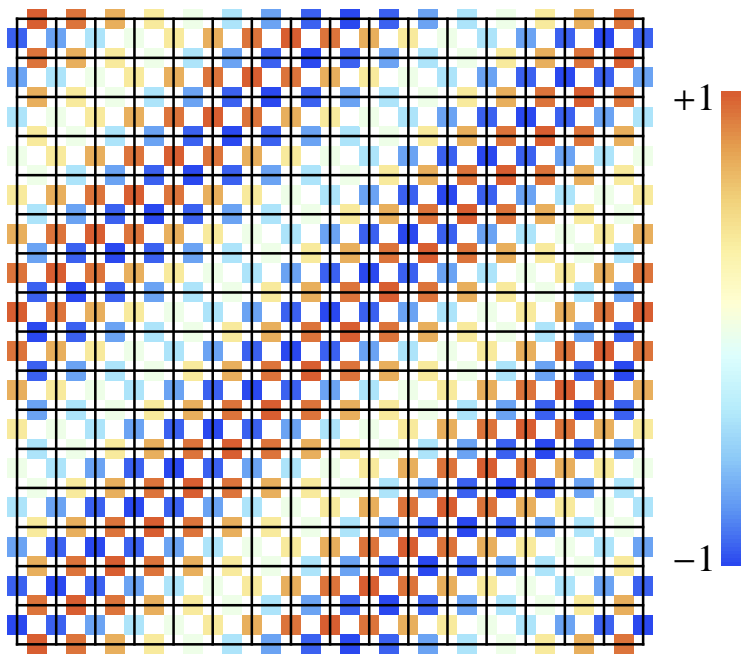}
\caption{Plot of the bond density modulations in (\ref{CN3}). 
The lines are the links of the underlying
square lattice. Each link contains a colored square representing the value of $\left \langle c^{\dagger}_{\vec{r} \sigma}  c_{\vec{s} \sigma} \right\rangle$, where $\vec{r}$ and $\vec{s}$ are the sites at the ends of the link.
We chose the ordering wavevector $\vec{Q}_1 = (2 \pi/16) (1,-1)$. Notice the local Ising-nematic ordering,
and the longer wavelength sinusoidal envelope along the diagonal.}
\label{fig:density}
\end{center}
\end{figure}

\begin{figure}
\begin{center}
\includegraphics[width=4in]{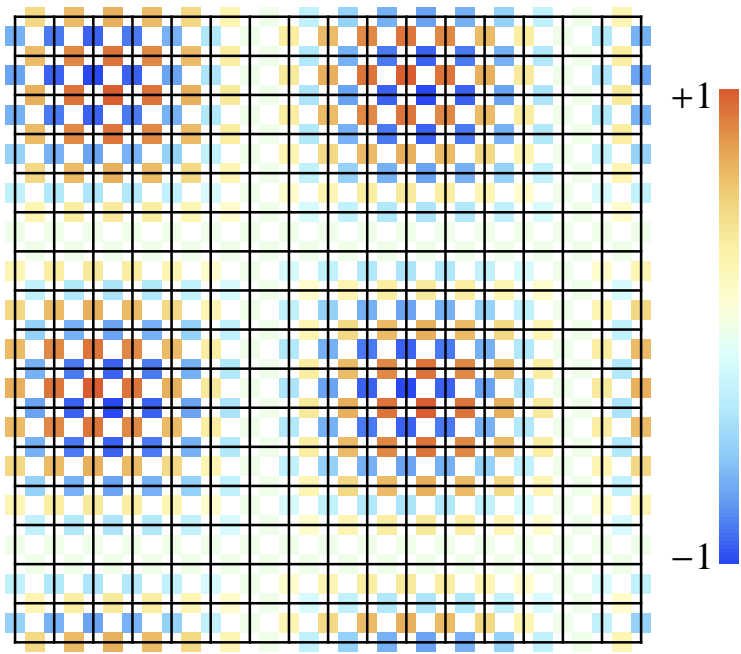}
\caption{As in Fig.~\ref{fig:density}, but for orderings along both
$\vec{Q}_1 = (2 \pi/16) (1,-1)$ and $\vec{Q}_2 = (2 \pi/16) (1,1)$.}
\label{fig:density2}
\end{center}
\end{figure}

What type of ordering does the susceptibility in Fig.~\ref{fig:nematic} correspond to?
We can deduce this by performing the pseudospin transformation on Fig.~\ref{fig:hotspots2}: this
changes the $\ell=3$ fermions from particles to holes, leading to
Fig.~\ref{fig:hotspots3}.
Using Eq.~(\ref{Cooperon}), this makes it clear that the scattering amplitude in Fig.~\ref{fig:nematic} is that associated with the
ordering of the operator
\beq
O =  \left( \psi^{1\dagger}_{1 \sigma} \psi^{3}_{1 \sigma}
- \psi^{1\dagger}_{2 \sigma} \psi^{3 }_{2 \sigma} \right), \label{do}
\eeq
and the sign between the two terms is the pseudospin descendant of the $d$-wave
pairing. Unlike, the Cooper pair operator, the operator $O$ carries a non-zero momentum
$\vec{Q}_1$, as shown in Fig.~\ref{fig:hotspots3}. For the cuprate Fermi surface, 
this is a small momentum oriented along a square lattice diagonal. This non-zero momentum would suggest that $O$ is a charge density wave at wavevector $\vec{Q}_1$.
However, this does not account for the relative sign in (\ref{do}), which implies that
$O$ changes sign upon reflection about the square lattice diagonal parallel to $\vec{Q}_1$.
The situation becomes clearer, when we write (\ref{do}) in terms of the underlying lattice
fermions $c_{\vec{k}\sigma}$. The state in which $\left\langle O \right\rangle \neq 0$
has
\beq \left\langle c^\dagger_{\vec{k} - \vec{Q}_1 /2,\sigma} c_{\vec{k} + \vec{Q}_1 /2} \right\rangle
\sim \left(\cos k_x - \cos k_y \right) \label{CN2} \eeq
Despite the $d$-wave-like structure, 
this order is {\em not\/} the popular $d$-density wave \cite{sudip}; the 
latter is odd
under time-reversal, and in the present notation takes the form
\begin{equation}
\left \langle c_{\vec{k} - \vec{Q}/2, \sigma}^\dagger c_{\vec{k} + \vec{Q} /2, \sigma} \right\rangle 
\sim i  \left( \sin k_x - \sin k_y \right), \label{dden}
\end{equation}
with $\vec{Q} = ( \pi, \pi)$. The order in (\ref{dden}) is not enhanced near the SDW critical point, while
that in (\ref{CN2}) is.
By taking the Fourier transform of (\ref{CN2}),
it is easy to see that $O$ does not lead to any modulations in the site charge density
$\left \langle c^{\dagger}_{\vec{r} \sigma}  c_{\vec{r} \sigma} \right\rangle$, and so it is not 
a charge density wave.
The non-zero modulations occur in the off-site correlations $\left \langle c^{\dagger}_{\vec{r} \sigma}  c_{\vec{s} \sigma} \right\rangle$ with $\vec{r} \neq \vec{s}$. For $\vec{r}$ and $\vec{s}$ nearest-neighbors, we have
\begin{equation}
\left \langle c^{\dagger}_{\vec{r} \sigma}  c_{\vec{s} \sigma} \right\rangle \sim 
\left( \langle O \rangle e^{ i \vec{Q}_1 \cdot (\vec{r} + \vec{s})/2} + \mbox{c.c.} \right)
\left[ \delta_{\vec{r} - \vec{s}, \hat{x}} + \delta_{\vec{s} - \vec{r}, \hat{x}} - \delta_{\vec{r} - \vec{s}, \hat{y}} - \delta_{\vec{s} - \vec{r}, \hat{y}} \right], \label{CN3}
\end{equation}
where $\hat{x}$ and $\hat{y}$ are unit vectors corresponding to the sides of the square lattice unit cell.
The modulations in the nearest neighbor bond variables $\left \langle c^{\dagger}_{\vec{r} \sigma}  c_{\vec{r}+\hat{x}, \sigma} \right\rangle$ and $\left \langle c^{\dagger}_{\vec{r} \sigma}  c_{\vec{r}+\hat{y}, \sigma} \right\rangle$ are plotted in Figs.~\ref{fig:density} and~\ref{fig:density2}. 
These observables measure spin-singlet correlations
across a link: if there are 2 electrons on the 2 sites of a link, this observable takes different values
depending upon whether the electrons are in a spin singlet or a spin triplet state. Thus $O$ has
the local character of a valence bond solid (VBS) order parameter. The first factor on the rhs of Eq.~(\ref{CN3})
shows that the VBS order has modulations at the wavevector $\vec{Q}_1$ along a square
lattice diagonal. As we saw in Fig.~\ref{fig:hotspots3}, $|\vec{Q}_1|$ is small and
so the first factor
in (\ref{CN3}) contributes a relatively long-wavelength modulation, as is evident from Figs.~\ref{fig:density} and~\ref{fig:density2}. 
This long-wavelength modulation serves as an envelope to the oscillations given by
the second factor in (\ref{CN3}). The latter indicates that the bond order has opposite signs on the $x$ and $y$ directed bonds: this short distance behavior
corresponds locally
to an {\em Ising-nematic \/} order, which is also evident in Figs.~\ref{fig:density} and~\ref{fig:density2}.
The ordering in (\ref{CN3}) becomes global Ising-nematic order in the limit 
$\vec{Q}_1 \rightarrow 0$. It is interesting and significant that a $\vec{Q}_1=0$ Ising-nematic instability
appears in numerical weak-coupling renormalization group analyses \cite{halboth,yamase}.

\section{Conclusions}
The ordering found in Section~\ref{sec:nematic} has not (yet) been directly detected in any
experiments on the cuprates. However, it does have a natural connection to a variety
of recent experiments \cite{ando02,kohsaka07,hinkov08a,taill10b,lawler10}. 
The ordering is present only in bond observables (see Figs.~\ref{fig:density}
and~\ref{fig:density2}), and so relates to the bond-centered modulations seen
in scanning tunnelling microscopy. Also, we have so far performed only a linear
stability analysis near the SDW critical point. After accounting for non-linearities,
and the interplay with the superconducting order parameter, generically
$\vec{Q}_1$ will move away from the value determined by the location of the hot spots:
there is no symmetry which pins the value of $\vec{Q}_1$. In general, non-linearities
prefer commensurate wavevectors, and so a shift to $\vec{Q}_1=0$ is not unreasonable.
In that case, as we noted above, global Ising nematic order would appear.

Applying a magnetic field will strongly suppress the instability to $d$-wave superconductivity
noted in Section~\ref{sec:pairing}. However, the particle-hole instability of Section~\ref{sec:nematic}
occurs in the zero charge channel, and so should be insensitive to the applied field. Thus it is possible that
the ordering in (\ref{CN2}) becomes the dominant instability at large magnetic fields, and so plays
a role in the structure of the Fermi surface and the quantum oscillations. 

\subsection*{Acknowledgements}
This research was supported by the National Science Foundation under grant DMR-0757145, by the FQXi
foundation, and by a MURI grant from AFOSR.

\end{document}